\documentclass[journal,final]{IEEEtran}
\ifCLASSINFOpdf
  \usepackage[pdftex]{graphicx}
\else
  \usepackage[dvips]{graphicx}
\fi
\usepackage[tight,footnotesize]{subfigure}

\begin{document}
\title{Visualizing a Dusty Plasma Shock Wave via Interacting Multiple-Model Mode Probabilities}
\author{Neil~P.~Oxtoby, 
        Jason~F.~Ralph, 
        C\'eline~Durniak, 
        and Dmitry~Samsonov.%
\thanks{Department of Electrical Engineering and Electronics, 
the University of Liverpool, Brownlow Hill, Liverpool, L69 3GJ, United Kingdom.}%
}%
\markboth{IEEE TRANSACTIONS ON PLASMA SCIENCE,~Vol.~39, No.~11, November~2011}%
{Oxtoby \MakeLowercase{\textit{et al.}}: Visualizing a dusty plasma shock wave via interacting multiple-model mode probabilities}
\maketitle

\begin{abstract}
Particles in a dusty plasma crystal disturbed by a shock wave are tracked 
using a three-mode interacting multiple model approach. 
Color-coded mode probabilities are used to visualize the shock wave 
propagation through the crystal.
\end{abstract}

\section{Introduction}
\IEEEPARstart{M}{icron-sized} ``dust'' particles suspended in a plasma of neutral 
molecules, ions, and electrons can be used to model the microscopic kinematics of 
matter.  The dust becomes charged due to collisions with charge carriers in the 
plasma, resulting in the dust particles interacting via a screened Coulomb force, 
known as a Yukawa, or Debye-H\"uckel force~\cite{Konopka2000}.  
Dusty plasmas can be particularly useful for investigating phase transitions, 
where a weakly-confined dusty plasma crystal is melted by a shock wave, as considered 
in this work, and previously in~\cite{Samsonov2004,Durniak2010:IEEE}.  
For a dusty plasma monolayer, single particles can be observed using a digital camera.  
In the most basic sense, visualizing the dust kinematics consists of processing a 
sequence of images to reveal the time-varying position of each particle.  
However, the dust kinematics can be visualized another way --- using dynamic-state 
estimation.

Estimating the state of a moving object based on remote measurements is known as 
target tracking~\cite{BarShalom,JFRopaedia}.  
In target tracking, a recursive filter that finds the ``best estimate'' of the dynamic 
state at each point in time is combined with a decision process for associating 
measurement data to filter data.  
In this work, we consider the extended Kalman filter (EKF)~\cite{BarShalom}, 
which is perhaps the most widely used estimation algorithm for nonlinear 
systems~\cite{Julier2004}.  
The EKF represents a piecewise linearization of nonlinear dynamics about the current 
estimate.  It performs very well if the initial errors in the state estimate are 
small and the dynamics are approximately linear in the region of state space covered 
by the errors~\cite{JFRopaedia}.  

The EKF combines the measurement data with state prediction from a dynamical model 
to produce an estimate which aims to minimize the mean-square error (in 
target position/velocity/acceleration) at each point in time.  
The relative weights given to the measurement, and to the prediction, 
reflect confidence in the accuracy of each.  These selectable confidence levels 
facilitate tuning of the filter design to help optimize the tracker performance.  
For discussions of EKF design parameters, we refer the reader 
to~\cite{BarShalom,Oxtoby2011}.

Here we track the position, velocity and acceleration of 3000 simulated dusty plasma 
particles as an initial crystal lattice is melted by a shock wave that we have induced.  
This requires the use of three dynamical models, or modes, for predicting the dust 
dynamics: one each for the ``presence'', ``absence'', and ``aftermath'' of the 
shock wave.  
Running the corresponding three EKFs in parallel and combining the filtered states 
using an interacting multiple model (IMM)~\cite{BarShalom} automatically assigns  
probabilities to each mode, based on their likelihood of being correct.  
Here we use these IMM mode probabilities to visualize the dusty plasma 
kinematics.

\section{Results}
\IEEEPARstart{T}{racking} 3000 particles is intractable using a single 
EKF-based dynamic state estimator~\cite{Oxtoby2011}.  
Here we have used 3000 independent trackers, each tracking a single particle.  

\begin{figure*}[!t]
  {\centering \includegraphics[width=1.1\textwidth,clip=true,trim=28 12 -28 20]{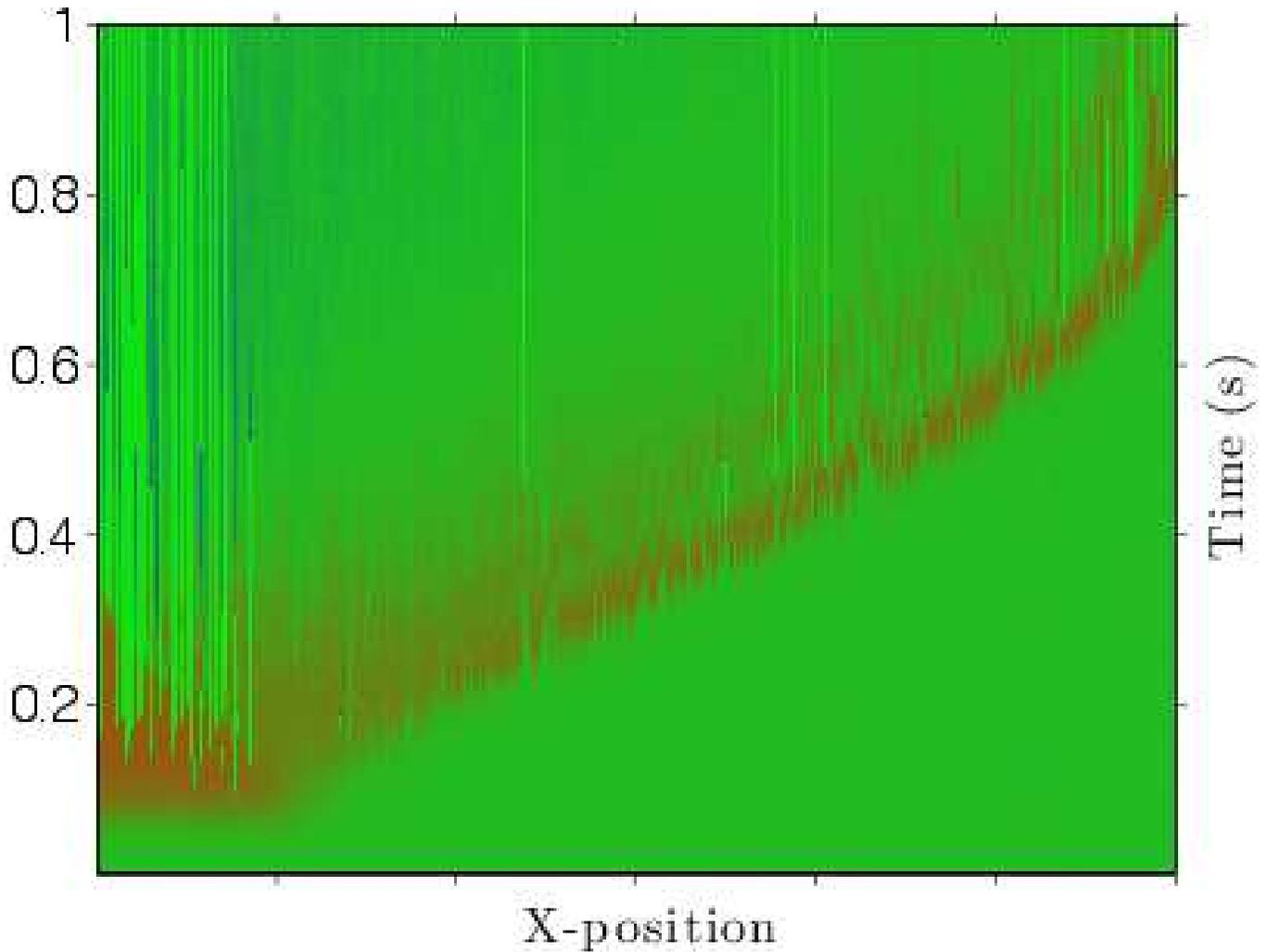}
  \caption{Color-coded image of IMM mode probabilities for X-position (horizontal axis) over time 
  (vertical axis).  
  The red/green/blue value for each pixel corresponds, respectively, to the mode probability in the 
  presence/absence/aftermath of the shock wave.  
  Propagation of the red shock front through the dust cloud is clearly visible, including 
  bifurcation after $t=0.3$s.  The feint blue upper-left corner shows post-shock-wave reflection 
  of particles off the dust cloud.}\label{fig_1}}
\end{figure*}

There are a number of selectable parameters involved in designing an extended Kalman 
filter, and an interacting multiple model tracker.  
In addition to careful mode design and noise-level selection, important design parameters 
for the IMM are the mode-transition probabilities, $p_{ij}$.  
These should reflect the actual system's mean sojourn times for each mode~\cite{BarShalom}.  
Here we have used 
\begin{equation}
  p_{ij} = \left[\begin{array}{ccc}
    0.70 & 0.05 & 0.25 \\
    0.05 & 0.70 & 0.25 \\
    0.05 & 0.05 & 0.90
  \end{array}\right] ,
  \label{eq:pij}
\end{equation}
reflecting the fact that we expect the particles to spend less time in modes 1 (``presence'') 
and 2 (``aftermath''), than in mode 3 (``absence'').  
The shockwave progression through the dusty plasma can be visualized via color-coded 
mode-probabilities, as shown in Figure~\ref{fig_1}.  
The red/blue/green intensity is determined by the probability for mode 1/2/3.  
The shock wave formation is clearly visible as a red region in the lower-left corner, 
with clearly visible shock-front propagation through space and time towards the 
upper-right corner.  Along the way, the shock front splits in two.  
Post-shock-wave reflection of dust particles off the dust cloud bulk is visible as 
the faint blue region in the upper-left corner of Figure~\ref{fig_1}.  

Figure~\ref{fig_1} shows that the IMM mode probabilities are useful for visualizing the 
dust kinematics in a dusty plasma.  Beyond visualization, specific physics about the dust 
can be extracted by an EKF-IMM tracker.  Examples include kinetic temperature (from particle 
velocities), particle-number density (from particle positions).  Indeed, any quantity 
related to the particle kinematics can be estimated in this manner.  
Importantly, in related work~\cite{Oxtoby2011}, we have found that a well-designed 
EKF-IMM tracker estimates the dust kinematics more accurately than particle tracking 
velocimetry (measurements alone).  
A future goal is to perform such dynamic estimation in real time, which would enhance the 
prospects for real-time control of a monolayer dusty plasma experiment.

\section{Conclusion}
Visualizing the kinematics of a monolayer of dust particles suspended in a 
plasma is not limited to using direct images of the dusty plasma.  
We have used an advanced dynamic-state estimation technique to track the 
kinetics of the dust particles (position/velocity/acceleration over time).  
Using a three-mode interacting multiple model state estimator based on 
an extended Kalman filter, we have produced particle-state estimates and 
associated mode probabilities.  
The mode probabilities reflecting the likelihood of each mode being 
a correct description of the dust particle dynamics.  
With a mode each designed for the presence, absence, and aftermath of 
a shock wave, we have visualized the dust kinematics (position vs.~time) 
by mapping the three mode probabilities to the red, green and blue color 
intensities in Figure~\ref{fig_1}.

\section*{Acknowledgment}

Financial support from UK EPSRC grant number EP/G007918 is acknowledged.  
N.P.O.~acknowledges use of high-throughput computational resources provided 
by the eScience team at the University of Liverpool.

\ifCLASSOPTIONcaptionsoff
  \newpage
\fi

\bibliographystyle{IEEEtran}
\bibliography{}

\end{document}